\documentclass[journal,a4paper]{IEEEtran}
\usepackage{graphicx} % for pdf, bitmapped graphics files
\usepackage{adjustbox}
\usepackage{amsmath}
\usepackage{float}
\usepackage{subfig}
\usepackage{caption}
\usepackage{tabularx}
\usepackage{pifont}
\usepackage{changepage}

\usepackage[boxed,ruled,vlined,linesnumbered,commentsnumbered, noresetcount]{algorithm2e}
\usepackage{algorithmic}

\usepackage{diagbox}
\usepackage{tikz}
\usetikzlibrary{shapes.geometric}
\usetikzlibrary{positioning}

\usepackage{tabularx}
\usepackage{ragged2e} % added for better adjust of cells' content
\newcolumntype{L}{>{\RaggedRight}X} % for cells with left aligned content
\usepackage{lipsum} % just for dummy text
\usepackage{amsmath}
\usepackage{amssymb}

\makeatletter
\newcommand\notsotiny{\@setfontsize\notsotiny\@vipt\@viipt}
\makeatother

\begin{document}

\title{Semantic Vehicle-to-Everything (V2X) Communications Towards 6G}

\author{
    \IEEEauthorblockN{Tengfei Lyu, Md. Noor-A-Rahim, Aisling O'Driscoll, Dirk Pesch}\\
    \IEEEauthorblockA{School of Computer Science and Information Technology, University College Cork, Ireland\\
    Email: t.lyu@cs.ucc.ie, md.noorarahim@ucc.ie, aisling.odriscoll@ucc.ie, dirk.pesch@ucc.ie}
}

%\author{Md. Noor-A-Rahim, Fadhil Firyaguna, Jobish John,   M. Omar Khyam,  Dirk Pesch~\IEEEmembership{Senior Member, IEEE,}  Eddie~Armstrong, Holger~Claussen,~\IEEEmembership{Fellow, IEEE,} and~H.~Vincent~Poor,~\IEEEmembership{Fellow, IEEE}

%\thanks{Md. Noor-A-Rahim, Fadhil Firyaguna, Jobish John,   Dirk Pesch,  and Holger Claussen  are with the  School of Computer Science \& IT, University College Cork,  Ireland.  (E-mail: {\tt \{m.rahim,ff28,j.john,d.pesch,h.claussen\}@cs.ucc.ie}). M. O. Khyam is  with the  Central Queensland University,  Australia. E. Armstrong is with the Johnson \& Johnson, Ireland (E-mail: {\tt earmstr1@its.jnj.com}). H. V. Poor is with the  Department of Electrical and Computer Engineering, Princeton University, USA (E-mail: {\tt poor@princeton.edu}).
%}}

\maketitle
%%\thispagestyle{empty}
%\pagestyle{empty}

%%%%%%%%%%%%%%%%%%%%%%%%%%%%%%%%%%%%%%%%%%%%%%%%%%%%%%%%%%%%%%%%%%%%%%%%%%%%%%%%
\begin{abstract}
 %This article presents a comprehensive study on integrating Semantic Communication (SEM-COM) within Vehicle-to-Everything (V2X) systems, a critical advancement as we progress toward the 6G frontier. It underscores the evolutionary trajectory of SEM-COM, marking a departure from traditional data-centric paradigms to an inherently intelligent and context-aware architecture. By elaborating on the refined design of network architectures conducive to this shift, this article presents an SEM-COM framework tailored to the intricate needs of modern vehicular communication systems. Through real-world use cases, this article articulates the potential of semantic V2X, showcasing how these concepts translate into tangible benefits for intelligent transportation. It further addresses the myriad challenges and contemplates future research directions, highlighting the necessity for ongoing innovation and exploration. The connections to 6G are woven throughout, delineating how the envisioned semantic capabilities are aligned with and instrumental for the next-generation wireless networks' goals of ultra-reliable, low-latency, and high-throughput communication services.

Semantic Communication (SEM-COM) has emerged as one of the disruptive technologies facilitating the evolution towards sixth-generation (6G) wireless networks. This article presents the potential of SEM-COM to transform Vehicle-to-Everything (V2X) communications, with a particular emphasis on its ability to enhance communication efficiency and intelligence. We discuss the core components and metrics that characterize SEM-COM, providing insights into its operational framework within the context of V2X communications. We illustrate the applicability and practicality of SEM-COM through real-world vehicular use cases, demonstrate the potential of SEM-COM to enhance aspects of intelligent mobility, such as communication efficiency and decision-making. 
%significant advancements toward intelligent mobility.
Finally, the article identifies key open research questions for SEM-COM V2X, pointing to areas that require further exploration and thus setting a foundation for future work in this evolving domain.

\end{abstract}


%\IEEEpeerreviewmaketitle

\section{Introduction}

Semantic communications, often referred to as SEM-COM, are one of the emerging tenets of 6G, aiming to provide a more intelligent and context-aware network. SEM-COM departs from the traditional data-centric paradigm by encompassing a broader interpretation of communication, including linguistic meanings and visual data. It proposes a shift from the conventional 'transmit first, then understand' approach to a more judicious 'understand first, then transmit' methodology. This signifies a dual shift — technical as well as conceptual — as it transitions away from Shannon's longstanding emphasis on data volume, focusing instead on the meaningfulness of the information exchanged~\cite{yang2022edge}.

It is envisioned that SEM-COM will be applicable to a broad spectrum of applications, ranging from augmented reality to autonomous vehicles and advanced healthcare. In the realm of vehicular networks,  SEM-COM could play a crucial role. This environment is highly dynamic, necessitating a communication system that can adapt rapidly to frequent changes in road conditions and vehicle speeds. The burgeoning number of vehicle sensors and advanced use cases requiring high reliability and low latency pose significant challenges for communications. These factors contribute to the scarcity of wireless resources, intensifying the need for efficient communication systems~\cite{noor20226g}. Moreover, privacy stands out as a critical issue in vehicular communications, Semantic communication enhances network security by transmitting only the necessary data while relying on the recipient's background knowledge to decode semantic information. This approach reduces the amount of data transmitted, thereby limiting the exposure of potential eavesdroppers. In addition, the semantic information extracted through encryption enhances the security of the transmitted data~\cite{basu2014preserving}. Given that V2X communication data may contain personal details such as license plates or images of individuals in urban settings, SEM-COM can potentially enhance privacy. For instance, instead of transmitting an image that could potentially identify a person, SEM-COM processes and transmits only essential semantic information, such as the presence of a person, which is sufficient for the context. This approach significantly reduces privacy risks by limiting the amount of sensitive data shared. %For instance, by processing and transmitting only the necessary semantic information rather than the full image, SEM-COM can mitigate privacy risks associated with sharing sensitive data.
%The more data transmitted, the greater the potential entry points for breaches

%Efficiency in V2X communication hinges on detailing only some environmental elements but on the timely and appropriate response to ever-changing scenarios. 
Efficiency in V2X communication does not just hinge on detailing only some environmental elements; it also depends on the timely and appropriate response to ever-changing scenarios. The utility of SEM-COM lies in its focus on the 'meaning' and 'context' of communication. 'Meaning' relates to the interpretation of content linked to the specific lexicon and semantics of the message, while 'context' involves the situational aspects of communication, such as the participating entities, their intentions, and the timing. These components are essential for understanding the message and ensuring that responses are both rapid and contextually appropriate~\cite{hosseini2022ccam}. SEM-COM is adept at transmitting fewer, yet more pertinent, data while facilitating improved decision-making, thus addressing the pressing demands of contemporary V2X communications, such as the need for high reliability, low latency, and enhanced security. SEM-COM is designed to enhance communication by effectively conveying a message's meaning and context, which could lead to clear and unequivocal communication—a critical need in the intricate domain of intelligent transportation systems. While promising, these capabilities are currently in developmental stages and represent aspirational goals~\cite{almeida2022mobile, 10038657}.
%confront challenges such as rapidly changing road conditions and the swift movement of vehicles, necessitating frequent and reliable data exchanges. These dynamic conditions exert considerable pressure on communication channels, which can struggle to satisfy safety and efficiency requirements, even at their maximum operational capacity.

%Efficiency in V2X communication hinges not on an exhaustive description of environmental elements but on the timely and appropriate response to dynamic conditions. The main concern is ensuring that the proper reactions to identified objects are executed correctly.Herein lies the utility of SEM-COM: 'meaning' relates to the interpretation of the content, often tied to the specific lexicon, symbols, and semantics of communication, while 'context' encompasses the situational aspects in which the communication takes place, including the participating entities, their intentions, and timing, all of which are crucial for understanding the message~\cite{hosseini2022ccam}.

The remainder of this article offers an in-depth exploration of SEM-COM within a V2X wireless network context. We provide an overview of semantic communications, highlighting the four foundational types and their potential to redefine conventional communication paradigms. Subsequently, we introduce the hierarchical network structure of SEM-COM V2X, which serves as a blueprint for integrating semantic layers into existing vehicular networks. Subsequent sections consider performance metrics like Age of Information (AoI) and its variants, which, while not exclusive to SEM-COM, are closely related and crucial for assessing the effectiveness of V2X communications. Following this, we outline four SEM-COM use cases, presenting their practical applications and showcasing practical applications and the benefits they bring to intelligent mobility. Finally, we present an analysis of the current challenges within this domain, pointing to future research and development needs in SEM-COM as we step towards the 6G frontier.

%The remainder of this article is organized to offer an in-depth exploration of SEM-COM within the V2X wireless network context. First we provide an extensive overview of semantic communications, highlighting its evolution as a transformative paradigm and its potential to transcend traditional network functions. We then delve into the specifics of semantic V2X communications, discussing network architecture, design considerations, and the incorporation of semantic principles into vehicular communication strategies. Subsequently, we examine the practical applications and implications of Semantic V2X in real-world scenarios, emphasizing its capacity to reshape vehicular communication systems and contribute to the advancement of intelligent transportation. We conclude by engaging with the challenges and prospective research avenues, setting the stage for continued exploration and innovation in Semantic V2X communications as we advance into the 6G landscape.

% igures/Type_of_SC_v1.5.png

\begin{figure*}[htbp]
\centering
\includegraphics{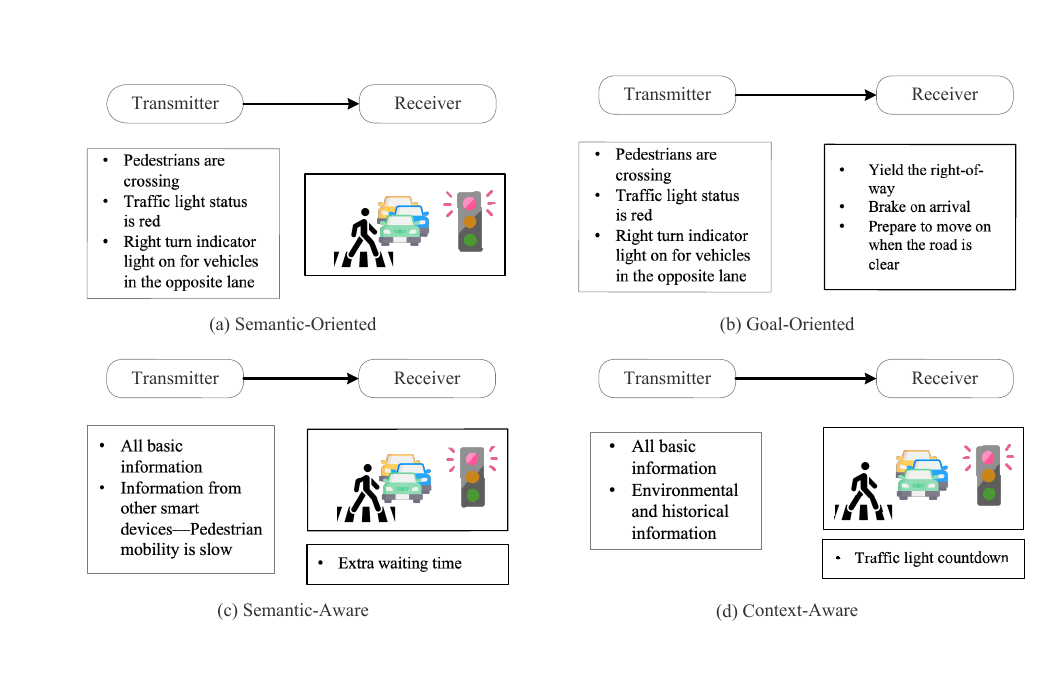}
%\caption{Illustration of data transmission and reconstruction by different types of SEM-COM when facing the same scenario.}
\caption{Illustration of data handling in different SEM-COM systems during the same traffic scenario}
\label{fig:Type_of_SEM-COM Systems}
\end{figure*}

\section{Overview of Semantic Communications}
%The emergence of Semantic Communication (SEM-COM) as a transformative paradigm promises to be the cornerstone of the sixth generation (6G) wireless network, providing intelligent, context-aware functionality that is critical to a wide range of applications. SEM-COM goes beyond the network's traditional role as a mere data conduit by combining the study of linguistic meaning with communication technology, upgrading the 'transmit-before-understanding' of information theory to an 'understand-before-transmitting' of semantic theory in order to enhance the interpretive power of the network~\cite{yang2022semantic, ch1971writings}. 

This section provides a brief overview of SEM-COM, initially focusing on its core components, followed by an analysis of the four distinct SEM-COM types. This demonstrates that the V2X example is used to illustrate semantic communication in general.
%Connected and Collaborative Automated Mobility (CCAM).

\subsection{SEM-COM Components}

%extract the parts of interest, known as semantic features. 
In the context of semantic communications, semantic features are the identifiable elements within data, while semantic information details how these features interact or relate. A key starting point is understanding its core components. State-of-the-art works offer various definitions, while combining in~\cite{yang2022semantic,chaccour2022less} offer varied definitions for the types of semantic communication, there is a consensus that some of the necessary architectural components, namely 'Semantic Encoder/Decoder' mechanisms and 'Background Knowledge.'

\begin{enumerate}
\renewcommand{\labelenumi}{\Alph{enumi}.}
    \item \textbf{Semantic Encoder:} Responsible for capturing the most information embedded in the source data and filtering out unnecessary redundancies. The focus is on accurately and efficiently encoding and transmitting the semantic information to ensure that the machine retains and understands its meaning, which is eventually extracted as semantic features. e.g., extracted from an image the current state of the light, filtering out background information such as the sky and pedestrians on the ground.

    \item \textbf{Semantic Decoder:} Inverse process of semantic encoding, where source data is recovered based on semantic information through various techniques. The focus is improving the receiver's ability to successfully infer semantic information from the communication data (semantic features). In some scenarios, semantic decoders may be unnecessary. e.g., in the multi-user collaborative vehicle ID retrieval scenario presented in~\cite{xu2023semantic}, the executor of the task can directly use the semantic features for ID retrieval without the need to recover the complete vehicle image, so the semantic decoder is omitted to improve the operational efficiency of the system.

    \item \textbf{Background Knowledge:} This is a pre-existing information base as well as a contextual understanding of the transmitter and receiver that is accurate in interpreting and processing semantic content. It ensures that communication exceeds technical accuracy and delves into contextual relevance and meaning. There are two prevailing implementations: one using knowledge graphs and the other using advanced artificial intelligence techniques~\cite{wheeler2023engineering}.
\end{enumerate}

\subsection{Type of SEM-COM}

Recent studies broadly categorize Semantic Communication (SEM-COM) into distinct types that address the varied demands of modern network systems. These include Semantic-Oriented communication, which focuses on the precision of semantic processing; Goal-Oriented communication, aimed at semantic accuracy and operational targets; Semantic-Aware communication, which augments task-oriented communication by improving shared knowledge; and Context-Aware communication, which integrates environmental understanding into strategy~\cite{yang2022semantic,chaccour2022less}.

% For a more graphic understanding, let us imagine a scenario where a vehicle at an intersection is blocked from view by a truck. In order to understand the situation at the intersection, and respond accordingly, the truck needs to transmit the data from the front camera to the vehicle, and the different types of SEM-COM systems will transmit different information. As shown in Fig.~\ref{fig:Type_of_SEM-COM Systems}. E2E stands for End to End, and ME2ME stands for Multi End to Multi End, i.e., there is more than one transmitter and receiver for Semantic-Aware and Context-Aware.

The core differences between them are illustrated in Fig.~\ref{fig:Type_of_SEM-COM Systems}, showing a common scenario. The view is obscured by a large truck in front of the vehicle. To safely navigate the junction, the obstructed vehicle requires data from the truck's front camera. The various SEM-COM systems, each with their unique approach, will communicate differing information relevant to this situation. The situation visible to the human eye includes several key details: the traffic light is red, indicating the need to stop; pedestrians are actively crossing the road, requiring drivers to yield; and the right turn indicator of the first vehicle in the opposite lane is on, signaling an intent to turn. For the vehicle behind the truck to safely cross the junction, the critical information it requires through SEM-COM systems includes the red light status to prompt slowing down, the presence of the pedestrian to signal, and waiting until the pedestrian has fully cleared the crossing. The activation of the turn indicator on the opposing vehicle suggests when it is safe to pass the junction.

\begin{enumerate}
\renewcommand{\labelenumi}{\Alph{enumi}.}
    \item \textbf{Semantic-Oriented:} This type prioritizes the accurate interpretation of significant observable elements, such as traffic lights and pedestrian movements, to ensure that the transmitted message maintains the original data's integrity. It focuses on the fidelity of the semantic content to ensure clear and precise communication.%Focuses on the significance of the observable elements, ensuring that the conveyed message maintains the original data's integrity. 

    \item \textbf{Goal-Oriented:} This system translates environmental observations into executable commands, aligning the vehicle's actions with situational demands. It ensures that the vehicle receives succinct instructions for immediate and appropriate responses, facilitating on-time execution of driving maneuvers.
    %Focuses on translating the observations into executable commands, ensuring the vehicle can complete the corresponding actions on time.

    \item \textbf{Semantic-Aware:} By integrating data from other smart devices at the intersection, this approach enhances decision-making processes. It accounts for the mobility of pedestrians and adjusts vehicular actions accordingly to maintain smooth traffic flow and heightened safety.
    %Data from other smart devices currently at the intersection will be incorporated to ensure that the vehicle is able to make decisions that enhance safety and traffic flow.

    \item \textbf{Context-Aware:} This type takes into account a broader set of data, including current traffic conditions, historical behavior patterns, and environmental factors. It synthesizes this information to develop comprehensive strategies for navigating intersections, enabling vehicles to make informed movements based on a thorough understanding of the environment.
    %More extensive data will be integrated, including current traffic conditions, other environmental factors, and historical data, to develop strategies for moving through the intersection.
\end{enumerate}

In light of the preceding exposition on SEM-COM types, it becomes clear that each approach carries its distinct set of challenges when applied within V2X environments. The 'Semantic-Oriented' approach, while prioritizing the fidelity of source data recovery, can lead to data redundancy and an increase in network load, which could be more practical for the bandwidth constraints of V2X applications. The 'Goal-Oriented' and 'Semantic-Aware' types, despite their focus on task relevance and agent cooperation, require constant adaptation and extensive customization that can pose real-time processing challenges in the dynamic vehicular context~\cite{yang2022semantic}. The effectiveness of Context-Aware's communication depends significantly on the quality and accuracy of the data it receives. Poor data quality can lead to incorrect interpretations and decisions~\cite{chaccour2022less}.

Based on the challenges inherent in these SEM-COMs, it becomes evident that their respective efficacies are markedly scenario-dependent within the V2X communication framework. The ’Semantic-Oriented‘ is especially effective in emergency response systems, where the precision of source data recovery is paramount. Conversely, the ’Goal-Oriented‘ is particularly suited to traffic management systems, where it integrates semantic accuracy with specific operational objectives. In collaborative settings like autonomous vehicle fleets, the ’Semantic-Aware‘ excels by augmenting the collective intelligence among vehicles, enhancing overall communication efficacy. Lastly, the ’Context-Aware‘ is optimally deployed in smart city applications, leveraging its adaptive communication capabilities to respond to dynamic environmental data. This contextual application of each SEM-COM type underscores the criticality of aligning the right technology with the specific requirements of diverse V2X scenarios.

\begin{figure*}[htbp]
\centering
\includegraphics{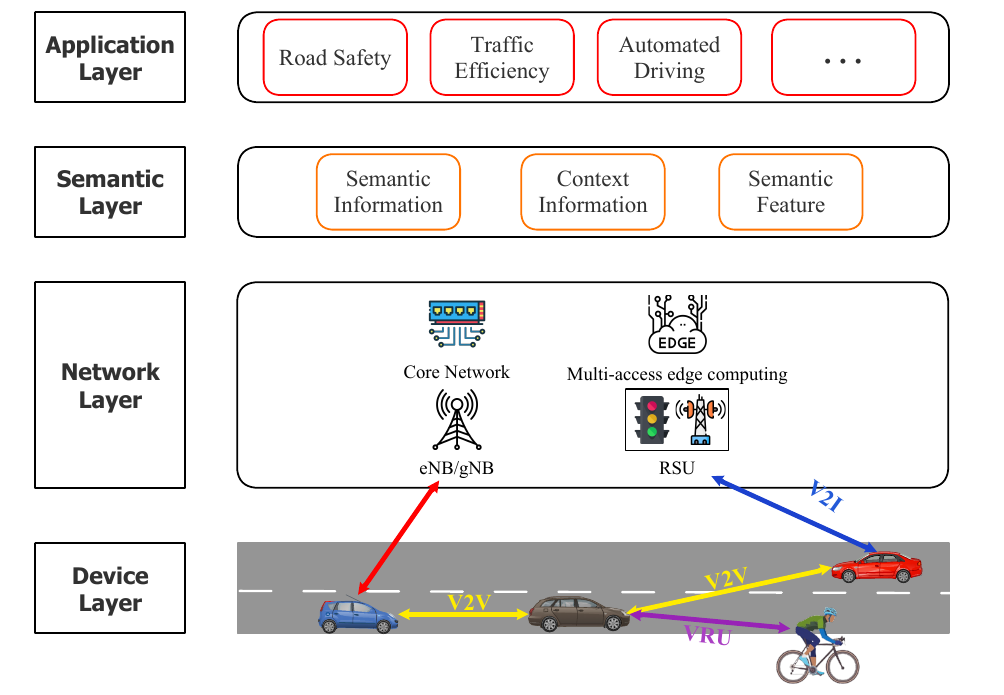}
\caption{Layered Overview of SEM-COM V2X System}
\label{fig:Network Architecture}
\end{figure*}

\section{Semantic V2X Communications}
This section discusses the integration of semantic communication technologies within V2X systems. A layered overview of the SEM-COM V2X communication system is provided, followed by a detailed exploration of the key metrics crucial for effectively implementing and assessing semantic communications in vehicular environments.

\subsection{Network Architecture and Design}
Despite the varying focuses and data utilization strategies of the four SEM-COM types previously discussed, they operate within a similarly structured network layering architecture, ensuring their foundational functionalities are aligned, as shown in Fig. ~\ref{fig:Network Architecture}. The Device Layer anchors this architecture, managing primary V2X communication streams such as vehicle-to-vehicle (V2V) and vehicle-to-infrastructure (V2I), among others, to maintain reliable and consistent operational data. Progressing to the Network Layer, the system's backbone emerges, equipped with multi-access edge computing (MEC) and roadside units (RSUs) designed for rapid and high-capacity data transmission, essential for real-time responsiveness in vehicular communications. Above this, the Semantic Layer processes semantic information and context-aware data, transforming raw inputs into actionable insights that enhance communication efficiency, as confirmed by the studies in~\cite{choi2022unified}. The architecture culminates with the Application Layer, which capitalizes on the semantically rich data to fuel various vehicular applications, enhancing road safety, improving traffic flow, and supporting autonomous driving initiatives. This topmost layer demonstrates the practical application of semantic processing, where complex data interpretations are effectively translated into tangible outcomes for the transportation ecosystem.
%The Device Layer is the bedrock, facilitating primary V2X communication streams: Vehicle-to-Vehicle (V2V), Vehicle-to-Infrastructure (V2I), Vehicle-to-Network (V2N), and Vehicle-to-Pedestrian (V2P). This layer ensures that the foundational exchange of operational data is consistent and reliable, setting the stage for advanced processing.

 %The Network Layer provides the essential backbone, composed of the core network, multi-access edge computing (MEC), and pivotal roadside units (RSUs). It is engineered to enable swift and high-capacity data transmission, a prerequisite for the real-time responsiveness required in vehicular communication systems.

 %Above this, the Semantic Layer processes Semantic Information, Context Information, and Semantic Features to bestow intelligent and context-aware communication capabilities. It transforms raw data into actionable insights, ensuring the efficiency and relevance of the information exchanged. Authors in~\cite{choi2022unified} confirm the feasibility of the semantic layer in real-world scenarios and demonstrate its effectiveness through probabilistic logic.

 %Finally, the Application Layer capitalizes on the semantic-rich data to drive various vehicular applications, from enhancing road safety to optimizing traffic flow and enabling autonomous driving. It represents the practical application of semantic processing, where the sophisticated interpretation of data is translated into tangible benefits for the transportation ecosystem.

 In exploring SEM-COM V2X, the research underscores a cooperative, semantic-aware architecture (Co-SC) that effectively conveys essential semantics from users to servers, significantly reducing data traffic while enhancing system performance in vehicular transportation systems. Studies offer a comprehensive survey of the evolution of V2X technologies, including early developments like dedicated short-range communications and the impact of big data and cloud-edge computing, highlighting the challenges and opportunities these technologies present for the Internet of Vehicles (IoV)~\cite{zhou2020evolutionary}. Big data's role within IoV is critically examined, emphasizing its importance in efficient processing, decision-making, and urban development while providing a taxonomy of big data usage in IoV and discussing the challenges and opportunities that arise~\cite{arooj2022big}. The integration of SEM-COM with V2X requires a careful and complex network architecture design for the IoV~\cite{xu2023semantic,zhou2020evolutionary,arooj2022big}, imposes the following requirements:
 
 %In exploring the SEM-COM V2X, Authors in~\cite{xu2023semantic} advocate for a cooperative, semantic-aware architecture (Co-SC) that efficiently conveys essential semantics from users to servers, significantly reducing data traffic while enhancing system performance in intelligent transportation systems~\cite{xu2023semantic}. Authors in~\cite{zhou2020evolutionary} provide a comprehensive survey of the evolution of V2X technologies, analyzing early developments like dedicated short-range communications (DSRC) and the impact of big data and cloud-edge computing, highlighting the challenges and opportunities these technologies present for the Internet of Vehicles (IoV)~\cite{zhou2020evolutionary}. Authors in~\cite{arooj2022big} emphasize the importance of big data in IoV, detailing its role in efficient processing, decision-making, and urban development while offering a taxonomy of big data usage in IoV and discussing key challenges and opportunities~\cite{arooj2022big}.

%In synthesizing the insights of Authors in~\cite{xu2023semantic}, Authors in~\cite{zhou2020evolutionary}, and Authors in~\cite{arooj2022big}, the integration of SEM-COM with V2X requires a careful and complex network architecture design for the IoV. This integration is essential to improve the system performance of ITSs and to utilize efficient, context-sensitive communications. SEM-COM meets these needs, but this integrated network architecture also imposes the following requirements:

\begin{enumerate}
    \item There is a need for high-speed data transmission to support real-time decision-making, dynamic context awareness to accommodate a variety of vehicle scenarios, and the scalability necessary to manage an expanding IoT network.

    \item The most important thing is to ensure that stable and reliable performance is provided in all environments and that efficient data management strategies are employed.

    \item Implement stringent security and privacy measures to protect the connected in-vehicle network.
\end{enumerate}

\subsection{Key Metrics}

% contextual information - what do others know and what information do I have that would allow me to interpret this
%age of information
%usefulness of information?
% Error metric in the context of SC
%\subsubsection{Contextual Information}
 %Contextual Information refers to understanding the background and environment relevant to a communication event, enhancing the interpretation and relevance of transmitted data within networks. This concept is fundamental in distinguishing SEM-COM from classical data-driven approaches. It ensures that the transmitted information is accompanied by sufficient context for the receiver to interpret it correctly. This is achieved by utilizing advanced artificial intelligence features that enable the receiver to not only decode but also understand and reason about the information it receives~\cite{chaccour2022less}. This means tailoring the data transmission to the known context of the receiver, thus improving the efficiency and effectiveness of the communication. In vehicular communications, Contextual Information enhances the V2X decision-making process by enabling vehicles to gain a deeper understanding of their surroundings, including road conditions, traffic patterns, and the intentions of nearby vehicles. This enhanced understanding is critical for making real-time decisions, ensuring road safety, and improving traffic flow. 

Key performance metrics crucial to the efficacy of information exchange within intelligent transportation systems.

\subsubsection{Age of Information}

The AoI measures the time that has elapsed since the generation of data at the source, representing the freshness of information in real-time communication systems~\cite{ayan2019age,beytur2020towards}.

\subsubsection{Age of Incorrect Information}

 The AoII quantifies the time since the last accurate and relevant update was received, indicating the prevalence of misinformation in the network~\cite{9796932}.

\subsubsection{Age of Outdated Information}

 The Ao2I gauges the duration since information at the destination became outdated relative to its source~\cite{9796932}. 

\subsubsection{Value of Information}

 The VoI assesses data's significance and decisional impact, playing a crucial role in optimizing system responses within vehicular communication networks~\cite{ayan2019age,molin2019scheduling}.
 
 %Value of Information (VoI), initially introduced in economics in 1966, has evolved into a key concept widely applied in the field of communications. Although interpretations may vary, a consensus on its core principles exists, focusing on enhancing network efficiency and information relevance.  VoI has found application in industrial automation and smart city management, playing a crucial role in reducing uncertainty in stochastic processes~\cite{ayan2019age,molin2019scheduling}. Additionally, VoI demonstrates adaptability by optimizing data sharing and state estimation in network management systems, and integrating with semantic communication for increased efficiency in cyber-physical systems and hierarchical control~\cite{uysal2022semantic}. In semantic V2X communications, VoI is expected to be vital for determining the importance and prioritization of data, particularly in the context of vehicular communications where timely and relevant data transfer is essential for safety and efficiency.

\begin{figure}
    \centering
    \includegraphics{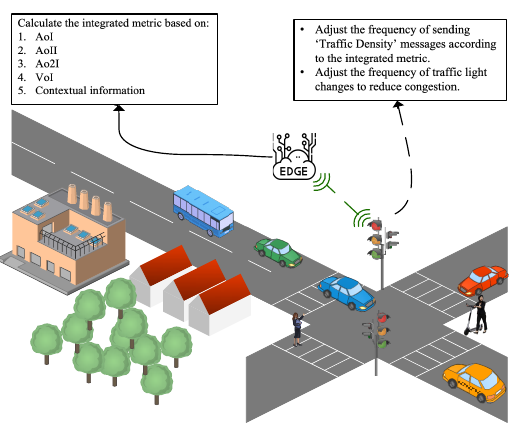}
    \caption{Adaptive Traffic Management Using SEM-COM}
    \label{fig:Integrated metrics}
\end{figure}

\textbf{Scenario Description:}: As shown in Fig.~\ref{fig:Integrated metrics}, an intelligent traffic light system powered by a semantic sensor measures the level of congestion at an intersection. The system employs edge computing to process this data along with one metric. Commands are then formulated based on this composite data, prompting the traffic light to dynamically adjust the color and duration of the lights to manage traffic flow efficiently. Moreover, the intelligent traffic light system adapts the frequency of congestion level reporting back to the edge computing system, depending on the real-time traffic conditions and the associated performance metric. The results of the above-mentioned metrics in adjusting the frequency of 'congestion level' sending in the face of this scenario are as shown in Table~\ref{Key Metric}.

\begin{table*}[htbp]
\caption{Congestion Level Information Reporting by the Traffic Light in Relation to Metrics~\cite{ayan2019age,beytur2020towards,molin2019scheduling,9796932}}
\label{Key Metric}
\centering
\scriptsize 
\renewcommand{\arraystretch}{1.2}
\begin{tabularx}{\dimexpr\textwidth-1.5cm}{l|X|X}
\hline\hline
\textbf{Metric}& \textbf{Congestion Level Information Reporting} &\textbf{Relevance and Utility in V2X (Out of 5)} \\
\hline
AoI& The traffic system regularly updates the 'congestion level' to provide current data for ongoing applications. & \ding{72} \ding{72} \ding{73} \ding{73} \ding{73} \\
%2/5 because while regular updates are good for real-time applications, the metric does not account for the dynamic nature of traffic conditions which can change very rapidly.\\
\hline
AoII& The system verifies and corrects 'congestion level' data accuracy to minimize misinformation and network traffic. & \ding{72} \ding{72} \ding{72} \ding{73} \ding{73} \\
%3/5 because it ensures data accuracy before transmission, reducing unnecessary network traffic and focusing on the correctness of information which is critical for decision-making.\\
\hline
Ao2I& Congestion updates are triggered based on data age to support timely decision-making in traffic management. & \ding{72} \ding{72} \ding{73} \ding{73} \ding{73} \\
%2/5 because it dispatches updates based on the age of information. While this ensures that outdated information prompts an update, it does not necessarily prioritize the accuracy or the actual value of the information.\\
\hline
VoI& Update frequency for 'congestion level' is adjusted based on critical need, enhancing V2X communication during peak times. & \ding{72} \ding{72} \ding{72} \ding{72} \ding{73} \\
%4/5 as it takes into account the significance and decisional impact of data, particularly during critical times such as rush hours or accidents, which is highly pertinent in V2X communications for efficient and safe traffic management.\\
\hline\hline
\end{tabularx}
\end{table*}

In Semantic V2X communications, the synthesis of critical metrics attains a heightened level of sophistication when contextual information is intricately woven into the evaluation framework. The triadic relationship between the Age of Information (AoI), the Age of Incorrect Information (AoII), and the Age of Outdated Information (Ao2I) is augmented by the contextual relevance of updates, yielding a nuanced understanding of the data's currency and its pertinence to specific vehicular scenarios. Alongside the Value of Information (VoI), which appraises the data's decisional impact, these metrics collectively establish a multi-dimensional lattice for assessing communication efficacy. By integrating contextual information, this lattice transcends mere temporal metrics, encapsulating contemporary vehicular networks' dynamism and complex exigencies. This integration ensures a sophisticated orchestration of data flow, optimally tuned to the exigencies of safety, operational efficiency, and the overarching demands of SEM-COM V2X communications.

\section{Vehicular Use-cases} \label{sec:UseCases}
%The emergence of V2X communications heralds a significant leap forward in automotive technology, promising to improve transport systems' safety, efficiency, and interactivity. At the heart of this paradigm shift is the integration of SEM-COM, a breakthrough approach that goes beyond traditional methods of data exchange. SEM-COM emphasizes the importance of context, meaning, and relevance of information shared between vehicles and their environments, as opposed to traditional communication technologies focusing on raw data transfer. 

This section examines four different vehicular use cases, each of which exemplifies the transformative impact of SEM-COM in addressing contemporary automotive challenges. These four use cases provide a comprehensive overview of the capabilities and benefits of SEM-COM in the automotive environment, highlighting its potential to redefine vehicular communication and contribute to the advancement of intelligent mobility solutions.

\begin{figure*}[htbp]
\centering
\includegraphics{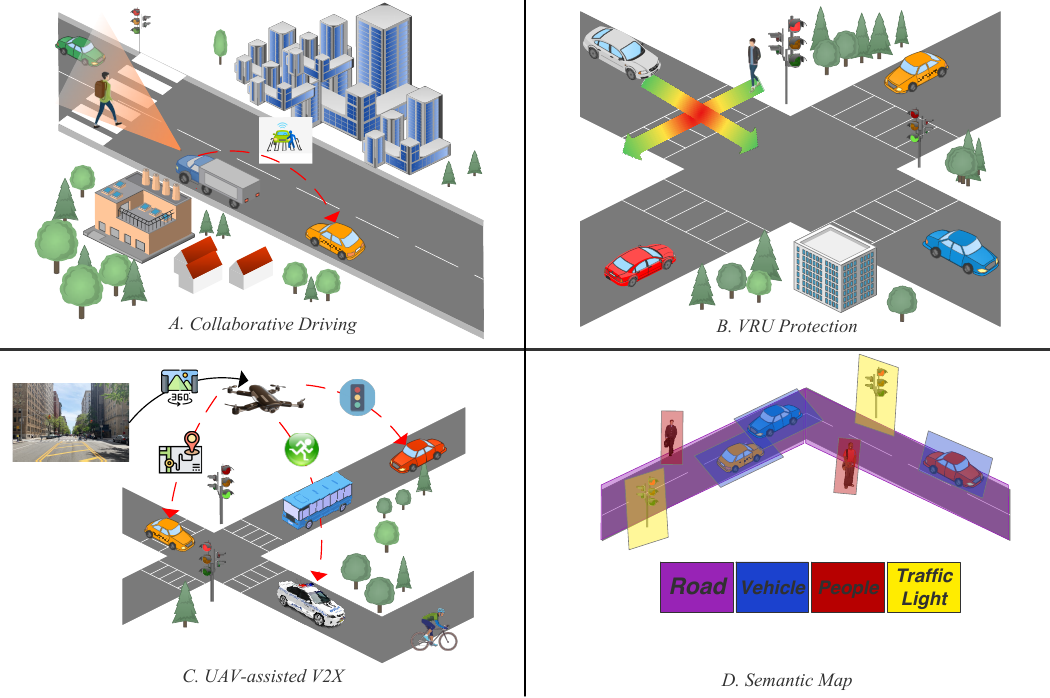}
\caption{V2X communication use cases demonstrates the applications of SEM- COM in Intelligent Transport System.}
\label{fig:Use cases}
\end{figure*}

\subsection{Collaborative Driving}
%e.g., see through another vehicle
\textbf{Scenario:} This scenario occurs when a vehicle's view is obstructed by another vehicle in front, necessitating insight into the road ahead beyond the immediate line of sight.

\textbf{Traditional V2X Approach:} Traditional V2X would transmit a real-time image from the front vehicle to the rear vehicle to provide visual information about the road ahead.

\textbf{SEM-COM Solution:} SEM-COM would transmit key semantic features (like distance to the next vehicle, speed, and traffic light status) from the front vehicle, enabling the rear vehicle to understand and react based on these specific, concise semantic cues.

\textbf{Comparison:} SEM-COM offers a more bandwidth-efficient and focused approach by transmitting only pertinent semantic features rather than entire images. This results in faster processing, less data transmission, and potentially quicker response times, which are crucial for safety in collaborative driving scenarios.

\subsection{VRU Awareness}

\textbf{Scenario:} This situation arises when a vehicle encounters pedestrians, necessitating either an alert to the pedestrians or an appropriate reaction from the vehicle to ensure safety.

\textbf{Traditional V2X Approach:} Traditionally, automated emergency braking (AEB) systems are employed in vehicles to prevent collisions with pedestrians. Audible beeps can be used to alert pedestrians in the vicinity of the vehicles.

\textbf{SEM-COM Solution:} SEM-COM can enhance pedestrian safety by transmitting detailed contextual information about the pedestrian's location, speed, and direction to the vehicle. The vehicle can then calculate the potential risk and either alert the pedestrian more effectively or adjust its own behavior (like slowing down or rerouting) based on a deeper understanding of the pedestrian's movements and intentions.

\textbf{Comparison:} While traditional systems rely on basic alerts or automatic braking, SEM-COM introduces a more proactive and nuanced approach. By understanding the context and predicting pedestrian behavior, SEM-COM can facilitate more tailored and potentially more effective responses, enhancing pedestrian safety and reducing the likelihood of accidents.

%\subsection{Platooning}

\subsection{UAV-assisted V2X}

\textbf{Scenario}: In this scenario, a cluster of UAVs captures high-definition images or videos from an overhead perspective to assist vehicles on the ground, e.g., in a crash or emergency situation.

\textbf{Traditional V2X Approach:} Traditionally, UAVs closest to the vehicle transmit these pictures or real-time videos directly to the vehicle, providing an aerial view of the surroundings.

\textbf{SEM-COM Solution:} In the SEM-COM approach, the UAVs would process these images or videos to extract meaningful contextual information. They would then send tailored semantic messages to the vehicle based on its specific context and requirements. This could include highlighting potential hazards, traffic conditions, or environmental changes that are relevant to the vehicle's journey. Furthermore, it is customized to send different relevant information to vehicles with different goals.

\textbf{Comparison:} Traditional methods provide a direct but potentially bandwidth-prohibitive stream of visual data, which the vehicle must then process. SEM-COM, on the other hand, offers a more efficient and focused approach by pre-processing this data into meaningful and actionable insights. This not only reduces the data processing load on the vehicle but also ensures that the vehicle receives the most relevant information in a timely manner, enhancing decision-making and situational awareness.

\subsection{Semantic Map}

\textbf{Scenario}: When a vehicle enters an unfamiliar road section, prompting a need to acquire a map to navigate the area securely and efficiently.

\textbf{Traditional V2X Approach:} Conventionally, vehicles would request high-precision maps from the cloud or roadside units to gain detailed information about the current road section's layout.

\textbf{SEM-COM Solution:}  The SEM-COM approach leverages cooperative efforts with other vehicles and roadside units, such as cameras, to construct semantic maps that encapsulate essential navigational data, optimizing travel through unfamiliar road sections.

\textbf{Comparison:} The SEM-COM methodology offers a bandwidth-efficient solution that dynamically updates semantic maps through real-time collaboration and information fusion. This enables vehicles to react promptly to real-time road conditions, ensuring safer and more informed navigation, particularly in areas where traditional mapping data may be insufficient or outdated.

\section{Challenges and Open Research Challenges}\label{sec:Chall}

\subsection{Background Knowledge}

Background knowledge is indispensable for accurately interpreting and transmitting context-rich information in vehicular environments. This knowledge base is primarily constructed through advanced artificial intelligence techniques bolstered by extensive datasets that inform and refine the training process~\cite{qin2021semantic}. Addressing the gap in semantic entropy akin to Shannon's concept of information entropy, yet without a validated formula for semantic entropy, is crucial for enhancing the predictability and optimization of vehicular communication systems. A direct correlation exists between the training model's expansiveness and the SEM-COM system's accuracy, necessitating larger models for more precise outcomes. However, integrating these sophisticated models into vehicle systems poses a challenge, especially in maintaining system responsiveness and computational efficiency. Furthermore, the dynamic nature of vehicular environments demands that SEM-COM systems exhibit a high degree of adaptability and be capable of contextual understanding and real-time communication adjustments in response to diverse environmental factors. Future research will aim to develop more sophisticated AI models that can dynamically adapt to the ever-changing contexts of vehicular environments while maintaining computational efficiency.

\subsection{Edge-based Computing}

Edge-based computing is pivotal for SEM-COM V2X communications, offering reduced latency and localized data processing critical for real-time vehicular decision-making~\cite{slamnik2021collaborative}. The implementation of edge computing in SEM-COM involves a seamless integration of vehicular networks with edge nodes, which must be robust and ubiquitous to ensure consistent performance. However, this integration introduces complexities in network architecture, requiring sophisticated coordination between vehicles and edge nodes. A significant challenge lies in balancing computational load distribution and maintaining rapid communication channels~\cite{qin2021semantic}. Moreover, vehicles' dynamic and mobile nature necessitates an edge computing infrastructure that can adapt to varying densities and speeds, ensuring uninterrupted and efficient communication flow. Further research is directed towards creating more resilient and adaptable network architectures. These should ensure uninterrupted communication and consistent performance across devices and conditions in vehicular networks.
%The following research stage will focus on creating more resilient and flexible network architectures that enhance the interplay between vehicular networks and edge nodes, ensuring consistent performance across diverse vehicular densities and speeds.

\subsection{Resource Management}
% The ever-increasing number of sensors enables connected cars to generate terabytes of data per hour of driving \cite{choi2016millimeter}.
Effective resource management is crucial in SEM-COM V2X communications, ensuring optimal utilization of computational and communication resources in vehicular networks~\cite{hwang2023resource}. This involves strategically allocating bandwidth and processing power, which is challenging given the fluctuating demands in a vehicular environment~\cite{choi2016millimeter}. Prioritizing data transmission based on urgency and relevance, particularly in SEM-COM, is a complex task requiring advanced algorithms. The challenge is further compounded by the need to balance resource distribution among a large number of vehicles and infrastructure elements. Additionally, ensuring energy efficiency in resource management while maintaining high performance is essential, especially in electric and hybrid vehicles where energy consumption impacts overall efficiency. Future investigations will center on devising advanced algorithms for real-time, adaptive resource allocation, ensuring optimal bandwidth and processing power distribution in the highly variable vehicular network environment.

%\subsection{Theoretical Limits}
%A fundamental challenge in advancing SEM-V2X communications is the nascent state of semantic information theory, which currently needs a comprehensive and universally recognized framework. The development of SEM-COM is intrinsically linked to the progress in this theoretical domain, where the absence of established principles limits the predictability and optimization of semantic communication systems~\cite{qin2021semantic}. This uncertainty in theoretical underpinnings poses a barrier to the standardization and widespread adoption of SEM-COM methodologies. Addressing this gap in semantic information theory is critical for unlocking the full potential of SEM-COM in vehicular networks and establishing reliable and efficient communication paradigms. Upcoming research will strive to develop a clear, universally recognized standard for semantic information theory, akin to Shannon's model, addressing the challenge of mathematically expressing semantic concepts for more effective device communication.

\subsection{Security and Privacy}
% Trade-off between semantic performance and security/privacy

In SEM-COM V2X communications, balancing enhanced semantic performance and robust security and privacy measures presents a significant challenge. Enhanced semantic capabilities often require the processing and transmitting of detailed and context-rich information, which can inadvertently expose sensitive data, raising privacy concerns~\cite{qin2021semantic,irshad2022security}. Conversely, stringent security protocols may impede the fluidity and richness of semantic communication, potentially diminishing its effectiveness. Developing SEM-COM systems that maintain high semantic performance while ensuring data security and user privacy is a complex endeavor, necessitating innovative approaches to encryption, data handling, and user consent mechanisms. This delicate balance is pivotal in fostering user trust and broad acceptance of SEM-COM V2X technologies. Future studies will explore innovative encryption methods and data handling protocols that enhance security and privacy in SEM-COM V2X communications without compromising semantic performance.

\subsection{Standardization}

Establishing comprehensive standards and protocols is paramount for ensuring interoperability among diverse systems and devices within the SEM-COM V2X ecosystem. The seamless integration and functioning of various components in this ecosystem hinge on universally accepted guidelines that facilitate communication and data exchange. However, the development of such standards for SEM-COM V2X is challenged by the evolving nature of semantic communication technologies and the need to accommodate a wide range of vehicular and infrastructural systems. Crafting these standards requires a concerted effort among industry stakeholders, regulatory bodies, and technology developers to create a unified framework that supports the diverse functionalities of SEM-COM V2X while fostering innovation and technological advancement. The absence of such standardization could lead to fragmentation and inefficiencies, impeding the widespread adoption and effectiveness of SEM-COM V2X communications. The forthcoming research effort will be directed toward developing unified standards and protocols that accommodate the evolving nature of SEM-COM technologies, ensuring interoperability and efficiency across diverse vehicular and infrastructural systems.

\section{Conclusion}\label{sec:Con}

SEM-COM is poised to be a formative technology in the growth of 6G networks and can address the bandwidth challenges posed by emerging vehicular applications. This article has outlined SEM-COM's key components, performance metrics, and network-layered architecture in the context of vehicular networks. Through real-world use cases, we have demonstrated SEM-COM's practical applications, showcasing significant advancements in intelligent mobility. Finally, we have pinpointed open research questions that signal a promising direction for future exploration, underlining SEM-COM's critical role in evolving intelligent and interconnected vehicular communications.

%In this article, we have embarked on an extensive examination of SEM-COM V2X communications, highlighting the transformative potential of SEM-COM within the domain of vehicular communication. Our journey commenced with a detailed overview of various SEM-COM systems, shedding light on their distinct functionalities and crucial role in augmenting vehicular interactions. As researchers continue to explore 6G networks, the significance of semantic communication becomes increasingly evident. It is poised to be a key component in the next generation of communication technologies, offering intelligent and efficient solutions for diverse future V2X applications. While SEM-COM is just one of several vehicular communication solutions, our discourse has illuminated its numerous advantages over traditional methods. By demonstrating the multifaceted benefits of SEM-COM, we underscore its potential to contribute substantially to advancing intelligent, interconnected, and environmentally conscious transportation systems.

\bibliography{references}
\bibliographystyle{IEEEtran}

\end{document}